\documentclass[aps,twocolumn]{revtex4}

\usepackage{graphicx}
\usepackage{comment}
\usepackage{cleveref}
\usepackage[usenames]{color}

\begin{document}

\bibliographystyle{apsrev}

\title{Detecting a Stochastic Gravitational Wave Background in the presence of a Galactic Foreground and Instrument Noise}
\author{\surname {Matthew} R. Adams and {Neil} J. Cornish}
\affiliation{Department of Physics, Montana State University, Bozeman, MT 59717}

\date{\today}

\begin{abstract}
Detecting a stochastic gravitational wave background requires that we first understand and model any astrophysical foregrounds.  In the millihertz frequency band, the predominate foreground signal will be from unresolved white dwarf binaries in the galaxy.  We build on our previous work to show that a stochastic gravitational wave background can be detected in the presence of both instrument noise and a galactic confusion foreground.  The key to our approach is accurately modeling the spectra for each of the various signal components.  We simulate data for a gigameter Laser Interferometer Space Antenna (LISA) operating in the mHz frequency band detector operating with both 6- and 4-links.  We obtain posterior distribution functions for the instrument noise parameters, the galaxy level and modulation parameters, and the stochastic background energy density.  We find that we are able to detect a scale-invariant stochastic background with energy density as low as $\Omega_{\rm gw} = 2 \times 10^{-13}$ for a 6-link interferometer and $\Omega_{\rm gw} = 5 \times 10^{-13}$ for a 4-link interferometer with one year of data.

\end{abstract}

\maketitle

\section{Introduction}

Primordial gravitational waves from the early Universe may be detectable by current and future gravitational wave observatories.  The Laser Interferometer Gravitational Wave Observatory (LIGO)~\cite{Abadie:2011fx} and Pulsar Timing Arrays (PTAs)~\cite{Jenet:2006sv} have already set bounds on the energy density in a stochastic gravitational wave background in their respective wavebands.  In this paper we show that complementary bounds can be set in the millihertz wave band with a space based interferometer.  To do so, we must properly account for instrument noise as well as any astrophysical foregrounds.  

Terrestrial detectors such as LIGO and PTAs have the benefit of being able to use cross-correlation between multiple detectors to separate stochastic signals from stochastic instrument
noise~\cite{Abbott:2009ws, Flanagan:1993ix, Allen:1996vm}.  With prospects for only one space based detector in the foreseeable future, we will not be able to employ the same technique.  In Ref.~\cite{Adams:2010vc}, hereafter Paper I, we showed that it is possible to separate instrument noise from a stochastic gravitational wave background by exploiting the differences in the transfer functions and spectral shape of the signal and noise contributions.

LIGO and PTAs also contend with astrophysical foreground signals.  In the very low frequency pulsar band, supermassive black hole binaries are expected to overwhelm any primordial stochastic signal~\cite{Jaffe:2002rt, Wyithe:2002ep, Enoki:2004ew, Grishchuk:2005qe, Jenet:2006sv}, and in the LIGO band, neutron star binaries could be a limiting factor~\cite{Regimbau:2001kx, Regimbau:2005ey, Abbott:2009ws}.  The main astrophysical stochastic background in the millihertz regime is expected to come from galactic white dwarf binaries~\cite{Bender:1997hs, Evans:1987qa, Schutz:1997bw}.  Many of these white dwarf binaries will be individually resolvable~\cite{Cornish:2007if}.   The rest will form a confusion foreground that could overwhelm any extragalactic stochastic signals if not properly modeled.  In this paper, we show that our ability to detect an isotropic stochastic gravitational wave background is not significantly reduced when the galactic foreground is properly modeled.  

In Paper I, we used simulated data for The NASA-ESA, Laser Interferometer Space Antenna (LISA) mission concept.  Simulated data was provided as part of the Mock LISA Data Challenges (MLDC) ~\cite{Babak:2009cj}.  While the original LISA partnership has dissolved,  ESA has a similar mission, eLISA~\cite{AmaroSeoane:2012je}, under consideration.  The main difference between the two missions is that the LISA concept has three interferometer arms whereas the eLISA concept has only two.  With 3 functioning arms, there are 6 laser links that exchange information between adjacent spacecraft.  For eLISA, there will only be 4 links.  We demonstrated in Paper I that while a 6-link configuration performs better, a 4-link configuration is still capable of separating a stochastic background from instrument noise.  

Here we extend the analysis of Paper I and show that both 6-link and 4-link detectors can detect a stochastic gravitational wave background in the presence of instrument noise and a galactic white dwarf foreground.  For comparison, we again use the LISA mission to demonstrate our technique.  We can remove one LISA arm as we did in Paper I, and it operates very much like eLISA.  The main difference is that eLISA will have shorter arm lengths than LISA and a diminished peak sensitivity that is shifted to slightly higher frequencies.

\begin{figure}[htbp]
   \centering
   \includegraphics[width=3.3in] {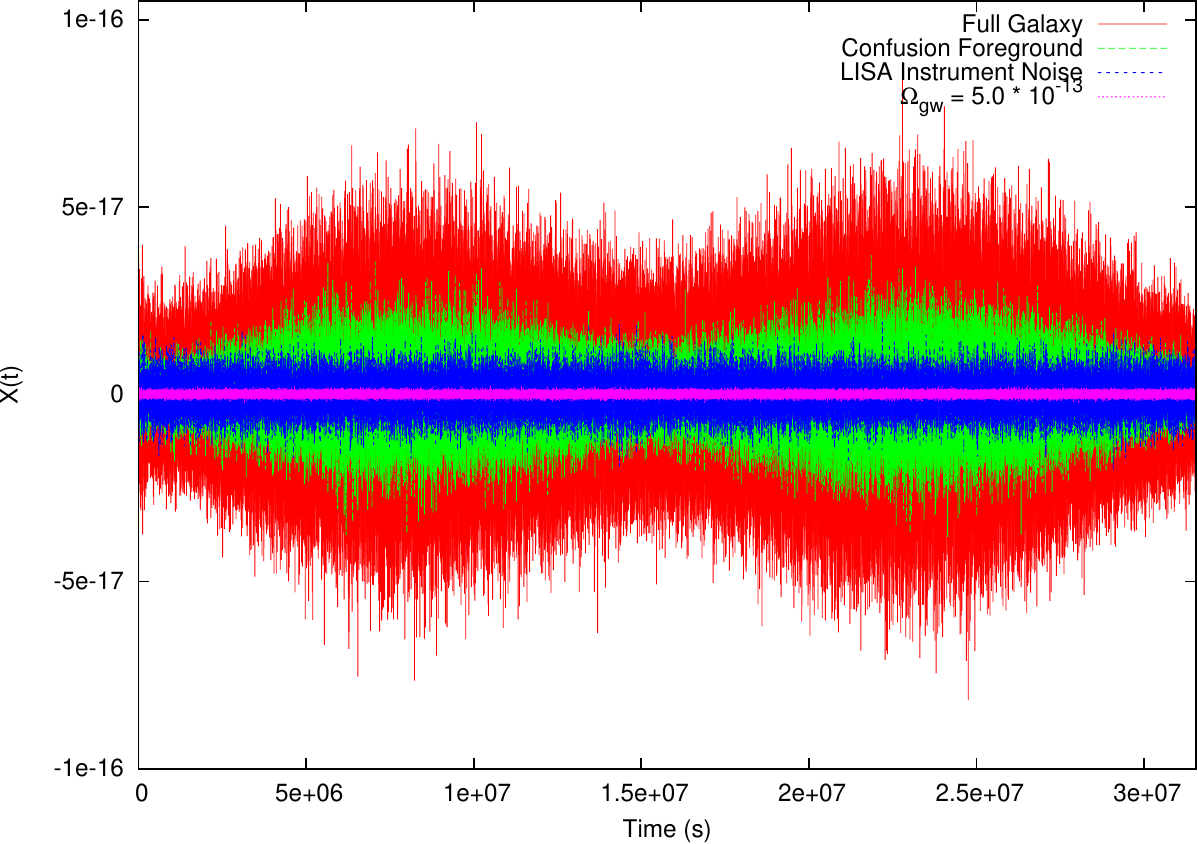} 
   \caption{The time domain noise, galaxy, and stochastic background signal components for the X-channel.  They have been bandpass filtered between 0.1 and 4 mHz.  We show that we are able to detect a scale invariant background with $\Omega_{gw} = 5 \cdot 10^{-13}$, which is well below the instrument noise and galaxy levels.}
   \label{fig:TD_modulation}
\end{figure}

There are two key factors involved in modeling the galactic foreground.  The first is the same idea used in Paper I.  The galactic foreground has a spectral shape that is distinct from the instrument noise and typical stochastic gravitational wave background models~\cite{Caprini:2009fx, Dufaux:2012rs, Binetruy:2012ze} (it would take an extremely fine tuned and bizarre primordial signal to match the spectral shape of the unresolved galactic foreground).  The differences in spectral shapes provide the main discriminating power amongst the three components.  In addition, most of the higher frequency white dwarf binaries will be individually resolved and regressed, meaning that the higher frequency data can be used to pin down parameters for the noise and the stochastic background with little or no galactic contamination. Secondly, the galactic foreground signal is modulated with a 1-year period due to the motion of the LISA constellation
around the Sun. As LISA cartwheels around the sun, the beam pattern will sweep across the sky.  The sweet spot of the beam pattern will hit different parts of the galaxy at different times throughout the year.  The variation in the detector response to the galaxy throughout the year creates the modulation in the signal.  Fig.~\ref{fig:TD_modulation} shows the full galaxy signal for 1 year of data.  The sweet spot of the beam pattern hits near the center of the galaxy twice throughout the year, giving the two peaks.  The figure also shows the confusion foreground after the bright binaries have been removed and the instrument noise level.  At certain times throughout the year, the galactic foreground will be much stronger than the instrument noise and a stochastic background.  It may seem that the galaxy could overwhelm any underlying signals, but as we found in Paper I, we can detect a background well below the instrument noise.  We will show in this paper that after properly modeling the galactic foreground, we are able to detect a background with energy density as small as $\Omega_{gw} = 5 \cdot 10^{-13}$ for a 4-link interferometer, as shown in Fig.~\ref{fig:TD_modulation}.

The amount of modulation in the signal depends on the shape of the galaxy and the distribution of white dwarf binaries throughout the galaxy.   Therefore, to accurately model the galaxy modulation, we need accurate measurements of the spatial distribution of white dwarf binaries in the galaxy.  One way to do this is to parametrize the galaxy shape distribution and simultaneously fit those parameters along with the noise and stochastic background parameters.  In principle this works, but we find that the confusion foreground signal can only provide very weak constraints on population model parameters.  In Ref.~\cite{Adams:2012qw}, hereafter Paper II, we showed that individually resolvable bright sources can be used to constrain the shape of the galaxy to levels better than current electromagnetic constraints.   We use the results from Paper II to constrain the galaxy distribution for the confusion foreground.  The amplitude of the bright binaries should also heavily constrain the amplitude of the confusion foreground.  We do not use the amplitude information here, but it is one potential way to improve our results.

The remainder of our paper is organized as follows:  In \S\ref{sec:Data}, we explain how we simulate the data for this project and compare to the MLDC data.  Before, we used low frequency approximation expressions for the transfer functions.  Here we use the full numerical expressions.
In \S\ref{sec:Model}, we explain how we model the noise, a stochastic background, and the confusion foreground.  We briefly review our analysis technique in \S\ref{sec:Analysis}.  Our results are presented in \S\ref{sec:Results} and in \S\ref{sec:Future} we discuss possible extensions to this work.

\section{Simulated Data}\label{sec:Data}

For this study, we simulated our own data and compare to the results used in Paper I where MLDC data was used.
We simulated our data for the galaxy using a code similar to the one used in the MLDCs~\cite{Cornish:2007if} and used our own codes for the noise and stochastic background.  We create a 1-year long data set sampled every 10 seconds.  We chose the low sampling rate for computational expediency.  The galaxy only extends to 3mHz and a stochastic gravitational wave background falls well below the noise shortly thereafter.  With a faster sampling rate, we could extend to higher frequencies.  We would better constrain the position noise, but our results would be otherwise unaffected.  

In comparison,  Paper I used the data from the third round of the MLDC.  Challenge 3.5 provides an approximately 3-week long data set of $2^{21}$ samples with 1 second sampling.  A scale invariant isotropic background was simulated with a frequency spectrum of $f^{-3}$.  The background was injected with a level that is $\sim 10$ times the nominal noise levels
at 1mHz, giving a range of $\Omega_{GW} = 8.95 \times 10^{-12} - 1.66 \times 10^{-11}$~\cite{Babak:2008aa} for the energy density in gravitational waves relative to the closure density per logarithmic frequency interval.  The MLDC noise levels were randomly chosen for each component with power levels that range within $\pm 20\%$ of their nominal values.  

We use the same noise model to create our noise realizations.  We simulate a stochastic background with the same spectral shape as the MLDC and use LISA simulator to generate a galaxy.  As the LISA spacecraft orbit, the distance between them fluctuates by approximately 3\%.  To simplify our analysis in Paper I we assumed an equal arm rigid LISA constellation.  This is not a bad approximation over short periods of time.  The MLDC data was less than a month long.  Here, our observation time is one year, but we break the year up into 50 segments.  Each segment is approximately one week in length.  The arm lengths will not change appreciably over that amount of time.  Additionally, the galactic foreground signal is also approximately constant over a one week period.  We generate the data in the frequency domain for each of the 50 segments.

\subsection{Instrument Noise}
We quoted expressions for the LISA noise spectral densities in Paper I.  Here we quote the noise amplitudes used to generated the noise spectra.  The three interferometric channels for LISA are generally referred to as X, Y, and Z.  The X-channel frequency spectrum is given by:

\begin{eqnarray}
X &=& 2i\sin\left(\frac{f}{f_*}\right)e^{f/f_*}\left[e^{f/f_*}(n^p_{13}-n^p_{12})+n^p_{31}-n^p_{21}\right]   \nonumber \\
  & & +4i\sin\left(\frac{2f}{f_*}\right)e^{2f/f_*}\Bigg[(n^a_{12}+n^a_{13}) \nonumber \\
	& & -(n^a_{21}+n^a_{31})\cos\left(\frac{f}{f_*}\right)\Bigg]
\end{eqnarray}
where $n_{ij}^p$ is the position noise for the link between spacecraft $i$ and $j$ and $n_{ij}^a$ is the acceleration noise.  The noise amplitudes are randomly drawn from strain spectral densities.
\begin{equation}
n_{ij}^B = \frac{\sqrt{S_{ij}}}{2}\delta
\end{equation}
where $\delta$ is a unit standard deviate and $B$ signifies either the real or imaginary part.  The strain spectral densities, $S_{ij}$, are randomly drawn to be within 20\% of the nominal values, which are:
\begin{equation}
S_p = 4 \cdot 10^{-42} Hz^{-1} \quad \quad S_a = 9 \cdot 10^{-50} s^{-4} Hz^{-1}
\end{equation}
for the position and acceleration noise respectively.

\subsection{Detector Response}
The details for how to calculate the detector response are given in Paper I as well as Ref.~\cite{Cornish:2002rt}.  We include here the main results needed to simulate our data.
The single arm detector response is given by:
\begin{equation}
D(\hat{k},f) = \frac{1}{2}(\hat{r_{ij}}\otimes\hat{r_{ij}}) \mathcal{T}(\hat{a}\cdot\hat{k},f)
\end{equation}
where $\hat{r}_{ij}$ is an arm vector and T is the single arm transfer function given by:
\begin{equation}
\mathcal{T} = \text{sinc}\left(\frac{f}{2f_*}\big(1-\hat{\Omega}\cdot \hat{r}_{ij}(t_i)\big)\right)\exp\left(i\frac{f}{2f_*}(1-\hat{\Omega}\cdot\hat{r}_{ij})\right)
\end{equation}
The signal is a convolution of the response with the gravitational waveform.
\begin{equation}
s(t) = D(f,k):h(f,t)
\end{equation}

In general, the gravitational waveform is given by a combination of the two polarizations:
\begin{equation}
h = h_+ \epsilon^+ + h_{\times} \epsilon^{\times}
\end{equation}
where $\epsilon^+$ are the basis tensors for the waves orientation with respect to the detector.  We can absorb the basis tensors into the detector response function to get the beam pattern functions:
\begin{equation}\label{beampattern}
F^P(\hat{\Omega},f) = \mathbf{D}(\hat{\Omega},f):\mathbf{e}^P(\hat{\Omega}) \, .
\end{equation}
We then rewrite the signal as:
\begin{equation}\label{signal}
s(t) = h_+ F^+ + h_{\times} F^{\times}
\end{equation}
Now $h_+$ and $h_{\times}$ depend only on the source parameters and we can plug in different wave templates.  We show below how $h_+$ and $h_{\times}$ are generated for the galaxy and stochastic background. 

\subsection{Galaxy}
The white dwarf waveforms~\cite{Timpano:2005gm} to be used in (\ref{signal}) are
\begin{eqnarray}\label{WD_waveform}
h_+(t) &=& A_+ \cos(2\psi)\cos(2\Omega t + \phi_o) \nonumber \\
            & &+ A_{\times} \sin(2\psi) \sin(2\Omega t + \phi_o) \nonumber \\
h_{\times}(t) &=& -A_+ \sin(2\psi)\cos(2\Omega t + \phi_o) \nonumber \\
                       & &+ A_{\times} \cos(2\psi) \sin(2\Omega t + \phi_o)
\end{eqnarray}
where the amplitudes are given by:
\begin{eqnarray}
A_+ &=& \frac{2G^2M_1M_2}{c^4r}\left(\frac{\Omega^2}{G(M_1+M_2)}\right)^{1/3}(1+\cos^2{\iota}) \nonumber \\
A_{\times} &=& \frac{4G^2M_1M_2}{c^4r}\left(\frac{\Omega^2}{G(M_1+M_2)}\right)^{1/3}\cos^2{\iota}
\end{eqnarray}
The phase is given by
\begin{equation}
\Phi(t) = 2\pi f_o + \pi \dot{f_o}t^2 + \phi_o - \Phi_D(t)
\end{equation}
where the modulation frequency $\Phi_D$ given by
\begin{equation}
\Phi_D(t) \equiv \frac{2\pi f_o}{c}\hat{k}\cdot \mathbf{x}_i(t)
\end{equation}
We use the catalog of white dwarf binaries developed by Gijs Nelemans containing approximately 29 million binaries~\cite{Arnaud:2007jy, Nelemans:2000es, Nelemans:2001nr}.   We give each binary a sky location drawn from the same spatial distribution used in Paper II:
\begin{equation}
\rho(x, y, z) = A e^{-r^2/R_b^2} + (1-A)e^{u/R_d} \text{sech}^2(z/Z_d)
\end{equation}
Here $R_b$ is a scale bulge radius, $R_d$ a scale disk width, $Z_d$ a scale height, $r=\sqrt{x^2+y^2+z^2}$, and $u=\sqrt{x^2+y^2}$.  The coefficient $A$ weights the number of stars in the bulge vs. the number in the disk.  We chose  $A=0.25$, $R_b = 500$, $R_d=2500$, and $Z_d=200$ as done in~\cite{Nelemans:2001hp}.  We calculate the SNR for each binary and designate sources with an SNR$>$7 as bright~\cite{Cornish:2007if}.  The bright sources are used in our analysis in Paper II and the remaining sources are used in the confusion foreground in this paper.

\subsection{Stochastic Gravitational Wave Background}

The stochastic gravitational wave background is generated using Eqn.~(\ref{beampattern}).  To simulate an isotropic background, we create equal area sky pixels using the HEALpix routines~\cite{Gorski:2004by}.  We generate $N=192$ sky pixels, the same number used in the MLDC.  In each sky pixel we randomly draw plus and cross polarization amplitudes for the stochastic background.\begin{equation}
h_A^B = \frac{\sqrt{\Omega_{gw}/N}}{2}\delta
\end{equation}
where A signifies the polarization plus or cross, B is the real or imaginary part, $\delta$ is a unit standard deviate, and $\Omega_{GW}$ is the energy density in gravitational waves per logarithmic frequency interval, scaled by the closure density.

\section{Complete Model}\label{sec:Model}

Our model is similar to the one used in Paper I.  We calculate strain spectral densities for the confusion foreground, the instrument noise, and a stochastic gravitational wave background for each of the interferometer channels (XX, AA, TT etc.).  We use an equal arm, stationary approximation for the LISA arm lengths.   The model is the sum of the three individual pieces:
\begin{equation}
XX = XX_{\text{noise}} + XX_{\text{sgwb}} + XX_{\text{galaxy}}
\end{equation}

\subsection{Noise Model}
We use the same noise model here that was used in Paper I.  Each spacecraft has two proof masses.  There will be a noise associated with travel between adjacent spacecraft in both directions.  For a 4-link mission, this gives a total of 8 noise parameters and for a 6-link mission there will be 12 noise parameters.  The X-channel position and acceleration noise contributions
are given by
\begin{equation} \label{eqn:XXp}
XX_p = 4 \sin^2 \left( \frac{f}{f_*} \right) \left( {{\it S_{12}^p}}+{{\it S_{21}^p}}+{{\it S_{13}^p}}+{{\it S_{31}^p}}\right)
\end{equation}
and
\begin{eqnarray} \label{eqn:XXa}
XX_a &=&16 \sin^2 \left( \frac{f}{f_*} \right) \bigg({{\it S_{12}^a}}+{{\it S_{13}^a}} \nonumber \\
           &  &+({{\it S_{21}^a}}+{{\it S_{31}^a}}) \cos^2 \left( \frac{f}{f_*} \right)  \bigg) \, .
\end{eqnarray}
Here $XX_p$ is the noise associated with position measurements of the proof masses, and $XX_a$ is the noise associated with the measurement of their accelerations.

\subsection{SGWB  Model}

In Paper I, we quoted a low frequency approximation to the beam pattern functions.  However, that is not what was actually used in the analysis.   As detailed in the paper, the injected stochastic background did not have the intended spectrum.  In our analysis, we used the MLDC training data to fit the spectrum of the background and used that approximate, numerical spectrum in our analysis.  Here we use the full equal-arm expressions for the LISA transfer functions.  Using Eqn.~(\ref{beampattern}), we calculate the detector response for each channel.
\begin{equation}
R_{ij}(f) = \sum_A \int \frac{d\Omega}{4\pi} F^A_i(\hat{\Omega},f) F^{A^*}_j(\hat{\Omega},f)
\end{equation}
The signal cross spectra are
\begin{equation}
\left<S_i(f),S_j(f)\right> = S_h(f)R_{ij}(f)
\end{equation}
with the gravitational wave spectral density given by
\begin{equation}\label{Sh_sgwb}
S_h(f) = \frac{3H_0^2}{4\pi^2}\frac{\Omega_{gw}(f)}{f^3}.
\end{equation}

As done in Paper I, we consider two cases, a flat spectrum where $\Omega_{GW}$ does not depend on the frequency, and a spectrum that allows for a power law dependence on frequency.  Since we won't know the slope of a stochastic gravitational wave background a priori, we want the flexibility in our model to fit for different slopes.
\begin{equation}
\Omega_{GW}^{slope} = \left(\frac{f}{1mHz}\right)^m
\end{equation}
In the latter case, Eqn.~(\ref{Sh_sgwb}) becomes
\begin{equation}
S_h(f) = \frac{3H_0^2}{4\pi^2}\frac{\Omega_{gw}}{f^{3+m/1mhz}}.
\end{equation}

\subsection{Galaxy Model}
There are two parts to our model for the galactic foreground.  We need a model for the spectral shape in the frequency domain, which is determined by the white dwarf binary population, and we need to model the modulation throughout the year, which is determined by the shape of the galaxy and LISA's orbital path.  We obtain the spectral shape by generating many different galaxies using the LISA Simulator.  We then smooth the spectrum in the frequency domain and average the spectra from each of the different realizations.  In practice, with a single galaxy observation, we could use the residuals from the bright source removal and the information they give about the galactic model to better constrain the spectrum.   

\begin{figure}[htbp]
   \centering
   \includegraphics[width=3.3in] {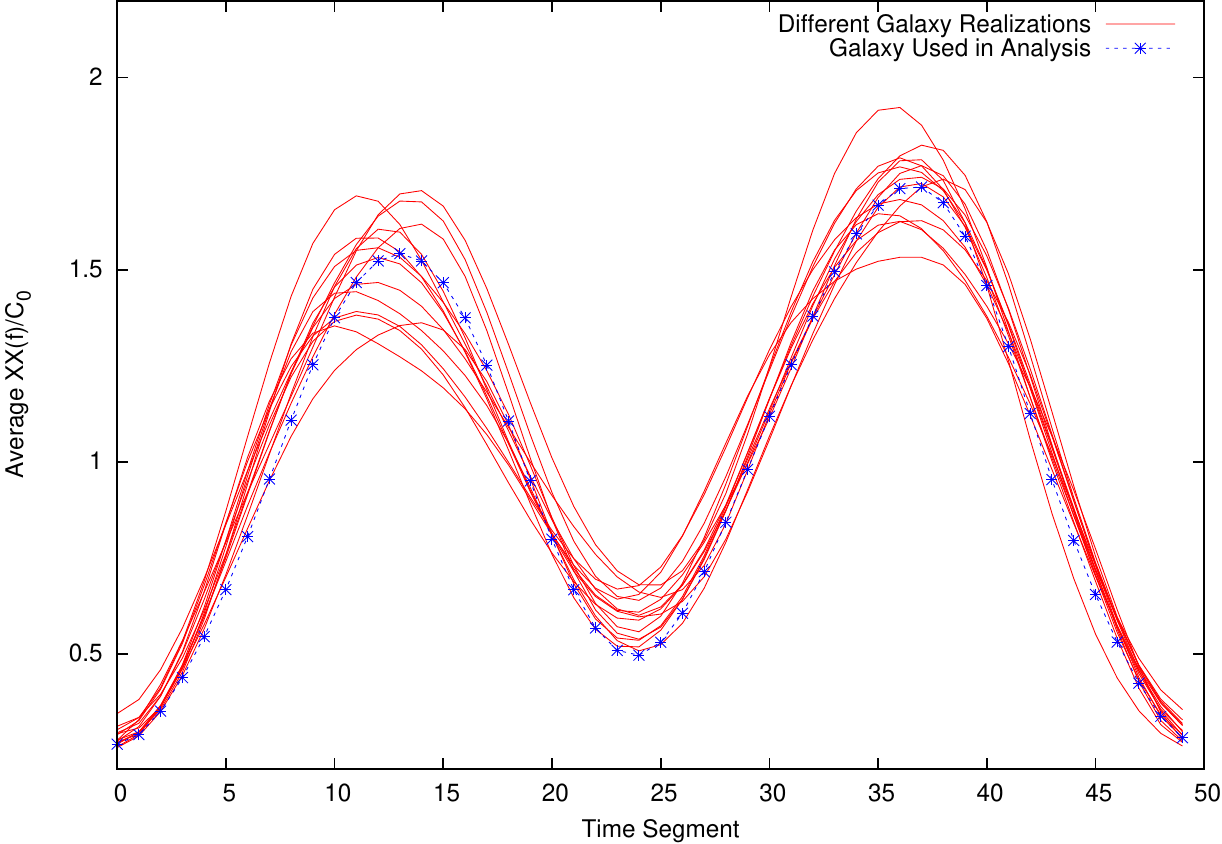} 
   \caption{Several galaxy realizations showing the scatter in the modulation levels throughout the year compared to the galaxy used in our simulations.  The points are scaled by the first Fourier coefficient, $C_0$.}
   \label{fig:modCurve}
\end{figure}

The modulation can be modeled by finding an average strain spectral density for each segment of the year.  For our approximately week long segments, the amplitude does not change appreciably and averaging the strain in each segment gives a good approximation to the modulation level for each week.  Fig.\ref{fig:modCurve} is made by plotting the average amplitude of each segment versus the central time for each weekly segment.  The two peaks correspond to the peaks shown in the time domain plot, Fig.~\ref{fig:TD_modulation}.  We see that the beam pattern slightly misses the center of the galaxy for the first peak, but hits it almost dead on for the second peak.  The amount of modulation, or difference between the peaks and troughs, depends on the shape of the galaxy.

If the only information we had about the galaxy came from the confusion foreground, we would need to include the galaxy distribution parameters in our analysis.  We can do this, but as mentioned before, the modulation of the confusion foreground does not sufficiently constrain the galaxy distribution parameters.  
Instead, we fix the galaxy parameters at the maximum a posteriori (MAP) values found by the analysis technique described in Paper II.  The bright sources constrain the shape of the galaxy so well, that including information from the confusion foreground would not tighten the posterior distributions of the galaxy shape parameters.  A more complete analysis would combine the bright source detection and
characterization with the stochastic background characterization in a simultaneous analysis of both components.

Even for the same galaxy parameters, different realizations of the galaxy will have significant variation due to the placement of a finite number of stars, as shown in Fig.~\ref{fig:modCurve}.  We need to account for the amount of variation that can occur from one realization to the next.  If we fixed the curve at the average value for the curves shown in Fig.~\ref{fig:modCurve}, the galaxy could take some power away from a stochastic background to try to fit the variation.  

\begin{figure}[htbp]
   \centering
   \includegraphics[width=3.3in] {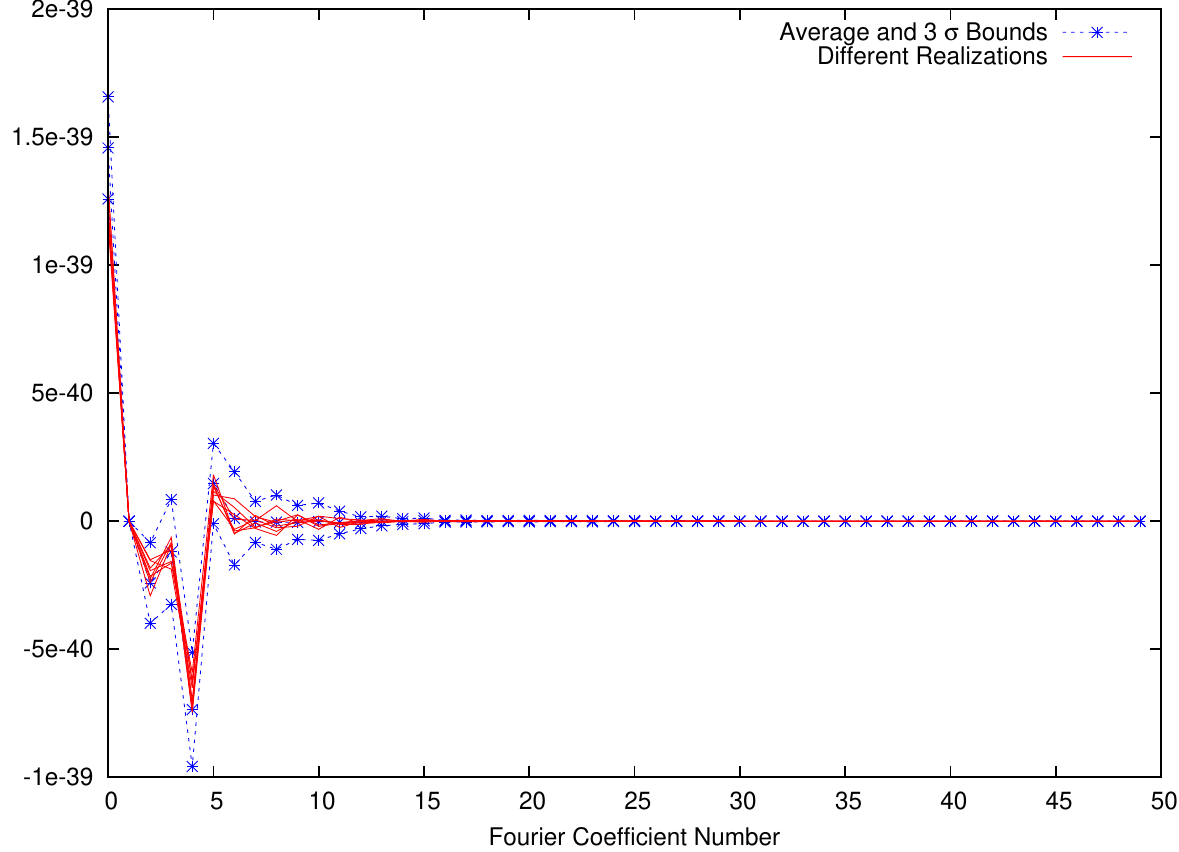} 
   \caption{Several galaxy realizations showing the scatter in the Fourier coefficients for different galaxy realizations.  The $\pm 3\sigma$ upper and lower bounds are used as the prior range on the Fourier coefficients in our analysis.}
   \label{fig:fourierCoeff}
\end{figure}

We could use the 50 amplitudes in the modulation curve as parameters in our model, but that is a large increase in our parameter space.
We instead use the Fourier coefficients as our model parameters.  The modulation curve in Fig.~\ref{fig:modCurve} can be uniquely characterized by 17 Fourier coefficients~\cite{Cornish:2002bh, Cornish:2001hg}.    Fig.~\ref{fig:fourierCoeff} shows the Fourier coefficients for several galaxy realizations using the same shape parameters.  To account for the variation from one simulation to the next, we simulate many galaxies with the same parameters.  We average the Fourier coefficients from the different runs and set the prior range to be $\pm 5\sigma$ around the average.  Fig.~\ref{fig:fourierCoeff} shows the average coefficients, the prior range, and the coefficient from several generated galaxies.  Using the Fourier coefficients reduces the number of galaxy parameters by more than a factor of two.  In practice, the savings are even better because only the first 5 or 6 coefficients are constrained very well.

\begin{figure}[htbp]
   \centering
   \includegraphics[width=3.3in] {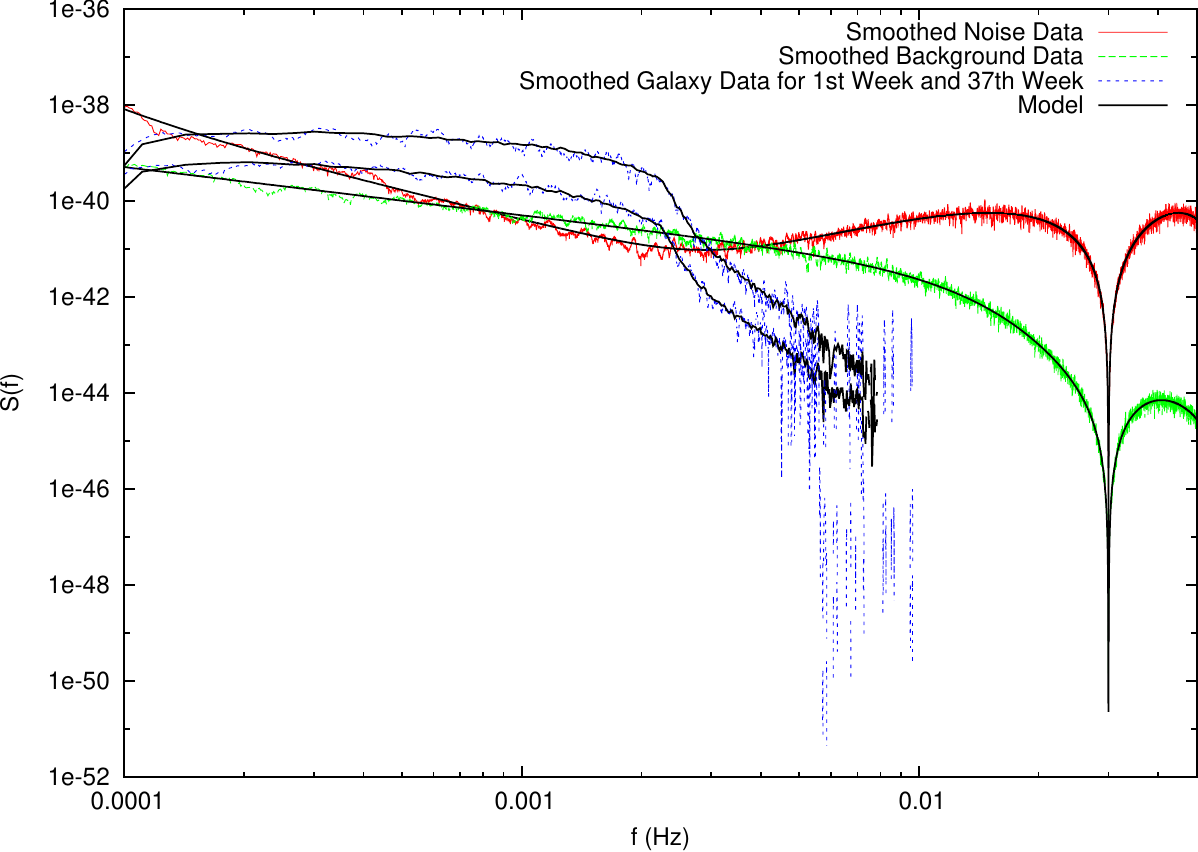} 
   \caption{Our smoothed, simulated data with the model overlaid in black (solid).}   \label{fig:modelComps}
\end{figure}

Fig.~\ref{fig:modelComps} shows smoothed data and our model for the various components.  The galaxy is shown during the first week of the year when the signal is at a minimum, and at a later time of year when the signal is at one of the peaks shown in Fig.~\ref{fig:modCurve}.  As mentioned earlier, we see the difference in the spectral shapes and that the galaxy extends over a shorter band.

\section{Analysis}\label{sec:Analysis}
We use the same analysis techniques used in Paper I to calculate posterior distribution functions (PDFs) for our model parameters and to do Bayesian model selection.  The likelihood used in Paper I is
\begin{equation}
p(\mathbf{X} | \vec{x}) = \prod \frac{1}{(2\pi)^{N/2}|C|}\exp\left(X_i C^{-1}_{ij} X_j\right)
\end{equation}
The only difference here arises from our treatment of the modulation.  We divide the year into 50 segments and calculate the likelihood for each segment.  We then take the product of the segments to get a total posterior distribution for all the data.
\begin{equation}
p(\mathbf{X} | \vec{x})= \prod \frac{1}{(2\pi)^{N/2}|C^d|}\exp\left(X_i^d {C^d_{ij}}^{-1} X_j^d\right)
\end{equation}
Here, d labels the time segments.

As in Paper I, we ran our analysis for both a full 6-link LISA as well as a 4-link version.  For the 6-link configuration, we use the orthogonal A, E, and T channels, which are combinations of the TDI variables X, Y, and Z~\cite{Prince:2002hp}.
\begin{eqnarray}
A &=& \frac{1}{3} (2X-Y-Z) \nonumber \\
E &=& \frac{1}{\sqrt{3}} (Z-Y) \nonumber \\
T &=& \frac{1}{3} (X+Y+Z)
\end{eqnarray}
We run over a frequency range of $0.8\text{ mHz} - 10\text{ mHz}$.  The low frequency cutoff is chosen to avoid some low frequency contamination in the generation of our data and stochastic background model.  At $10 \text{ mHz}$, the acceleration noise, galaxy, and stochastic background have all fallen below the position noise levels.  Any higher frequency bins would only help in further constraining the position noise.

\section{Results}\label{sec:Results}
\begin{figure}[htbp]
   \centering
   \includegraphics[width=3.3in] {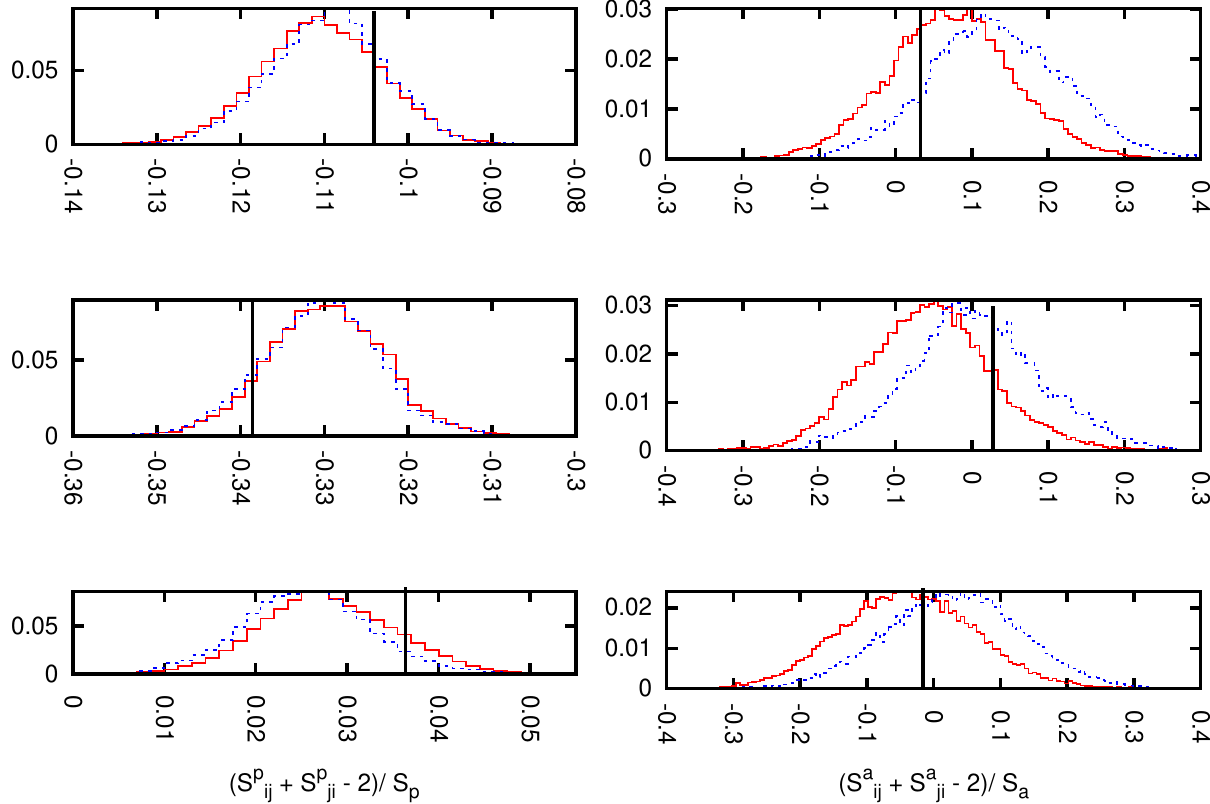} 
   \caption{The PDFs for the 6 constrained noise parameter combinations for the AET channels.  The position noise parameters are on the left and the acceleration noise parameters on the right.  The black (solid) vertical lines show the injected values.  The blue (dashed) PDFs included slope fitting and the red (solid) PDFs did not.}
   \label{fig:noise33_AET}
\end{figure}

\begin{figure}[htbp]
   \centering
   \includegraphics[width=3.3in] {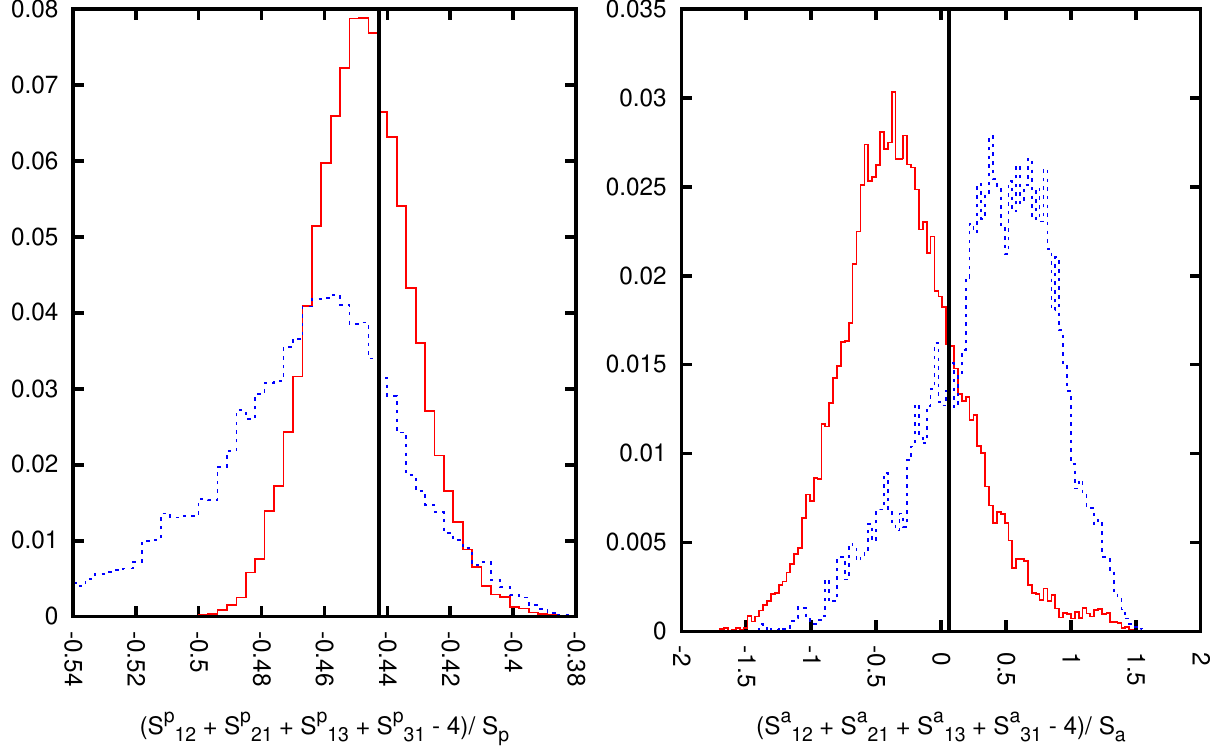} 
   \caption{The constrained position (left) and acceleration (right) noise parameters for X.  The black (solid) vertical lines show the injected values.  The blue (dashed) PDFs included slope fitting and the red (solid) PDFs did not.}
   \label{fig:noise24_X}
\end{figure}

Our analysis provides uncertainties for the instrument noise levels, the galaxy shape Fourier coefficients, and the stochastic background energy density and spectral slope.
Figs.~\ref{fig:noise33_AET} and~\ref{fig:noise24_X} show the posterior distributions for our noise model parameters.  As discussed in Paper I, only the total noise (i.e. $S_{12}^p + S_{21}^p$) in each interferometer arm is well constrained.  The individual noise levels are not individually constrained because only the distance between two adjacent spacecraft affects the interferometer output, not the individual movement of a single spacecraft. 

For the 4-link configuration, the sum of all four noise levels is constrained.  This can be seen from Eqns.~\ref{eqn:XXp} and~\ref{eqn:XXa}.  The position noise extends over a larger frequency band and is better constrained than the acceleration noise.

\begin{figure}[htbp]
   \centering
   \includegraphics[width=3.3in] {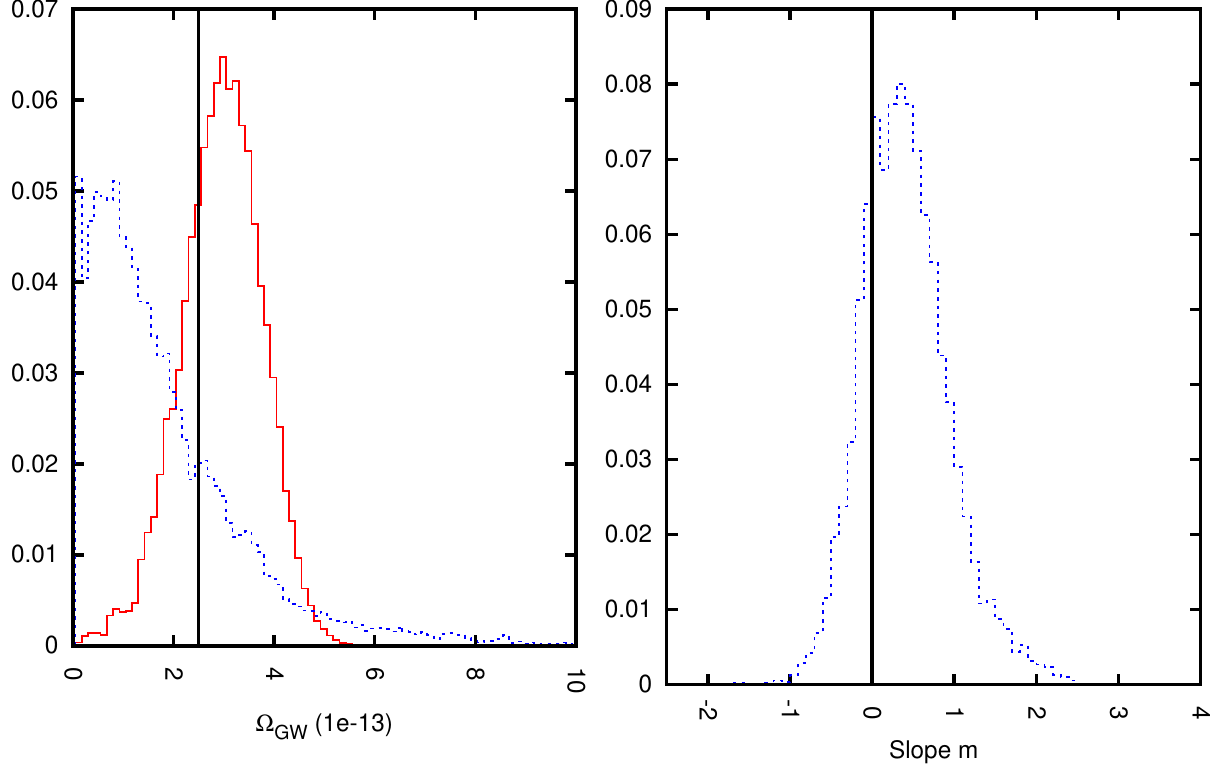} 
   \caption{The stochastic gravitational wave background level, $\Omega_{gw}$ and slope parameter $m$ for AET.  The black (solid) vertical lines show the injected values.  The blue (dashed) PDFs included slope fitting and the red (solid) PDFs did not.}
   \label{fig:sgwb33_AET}
\end{figure}

\begin{figure}[htbp]
   \centering
   \includegraphics[width=3.3in] {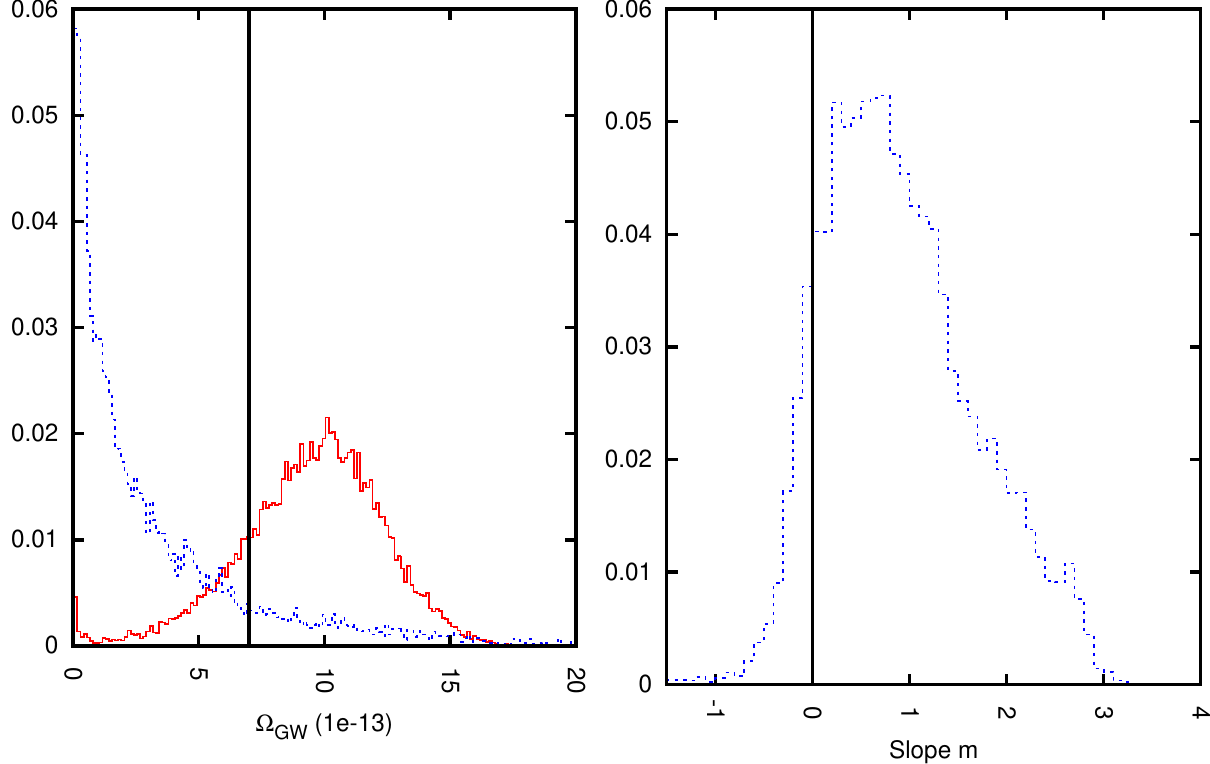} 
   \caption{The stochastic gravitational wave background level, $\Omega_{gw}$ and slope parameter $m$ for X.  The black (solid) vertical lines show the injected values.  The blue (dashed) PDF included slope fitting and the red (solid) PDF did not.}
   \label{fig:sgwb24_X}
\end{figure}

Figs.~\ref{fig:sgwb33_AET} and~\ref{fig:sgwb24_X} show the posterior distributions for $\Omega_{GW}$ for both the flat spectrum case and the frequency dependent case.  In Fig.~\ref{fig:sgwb33_AET}, the injected background level is $\Omega_{GW}=2.5\cdot 10^{-13}$ and in Fig.~\ref{fig:sgwb24_X} it is $\Omega_{GW}=7\cdot 10^{-13}$.  We show later that these are approximately the lowest background levels that could be confidently detected for the AET channels and X channel respectively.

When the proposed slope of the stochastic background matches or nearly matches the slope for either the instrument noise or the galactic foreground, there will be some correlation between the model parameters.  This leads to greater spreads in the PDFs of the model parameters.  Figs.~\ref{fig:sgwb33_AET} and ~\ref{fig:sgwb24_X} show that the PDFs from the model with a spectral slope are indeed broader than the PDFs for the model with $m=0$.  The effect is more pronounced for the X-channel case than for AET.  At these low detection levels, the slope would be uncertain over several integer values.  In Fig.~\ref{fig:bayesMLDC_NoGal} -~\ref{fig:bayesFull_LowDiv}, we see that this increase in uncertainty leads to a higher upper bound on the stochastic background level.

\begin{figure}[htbp]
   \centering
   \includegraphics[width=3.3in] {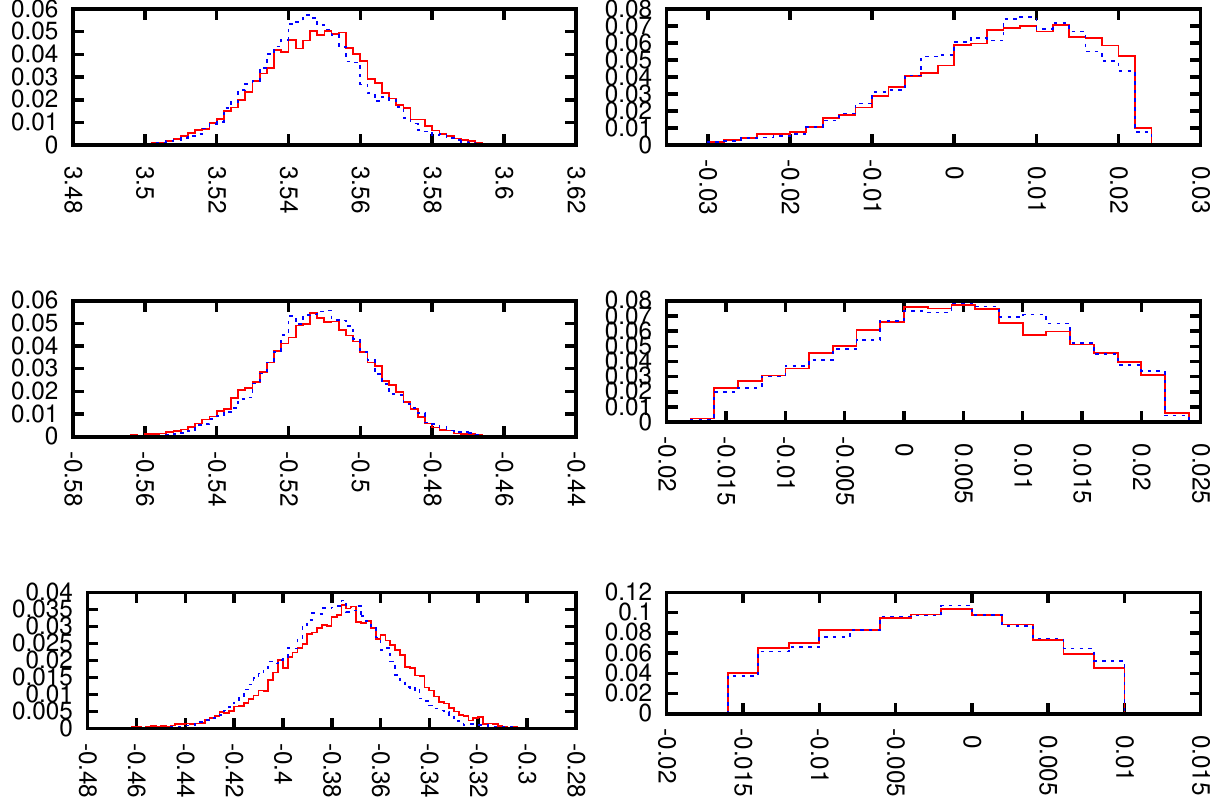} 
   \caption{The first three (left) and last three (right) Fourier coefficients for the A-channel.  The first coefficients are well constrained while the later ones are not.  The blue (dashed) PDFs included slope fitting and the red (solid) PDFs did not.}
   \label{fig:galaxyA_33_1}
\end{figure}

\begin{figure}[htbp]
   \centering
   \includegraphics[width=3.3in] {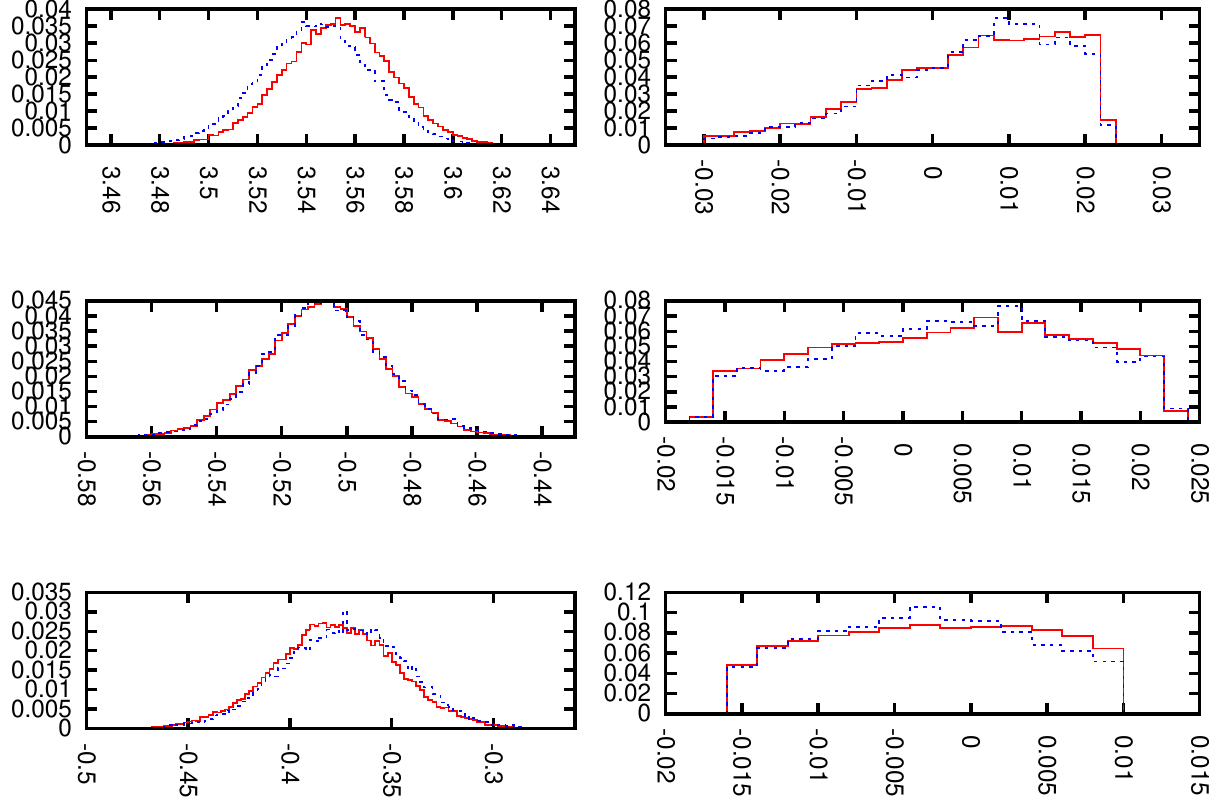} 
   \caption{The first three (left) and last three (right) Fourier coefficients for the X-channel.  The blue (dashed) PDFs included slope fitting and the red (solid) PDFs did not.}
   \label{fig:galaxyX_24_1}
\end{figure}

Figs.~\ref{fig:galaxyA_33_1} and~\ref{fig:galaxyX_24_1} show the PDFs for the galactic foreground model parameters.  We show the first three Fourier coefficients and the last three.  The $C_0$ coefficient sets the DC amplitude level for the galaxy.  The other Fourier coefficients determine the shape.  While there are 17 Fourier coefficients, the basic shape is determined by the first
5 or 6 and the higher coefficients only add fine details that are not well resolved in our analysis.  

\subsection{Bayesian Model Selection}

In Paper I, we suggested that the model with a spectral slope parameter should always perform worse than the model without a slope.  The extra degree of freedom will allow the model to fit the data better, however, it also comes with a penalty.  In Bayesian model selection, higher dimensional models have a larger prior volume to explore.  There is a penalty built in for having more degrees of freedom.  Since the injected data had a spectral slope of $m=0$, we wouldn't expect the model that allows for spectral slope fitting to ever outperform the model that assumes $m=0$.  

However, our results in Paper I showed that both models performed comparably well.  We postulated that this was do to our having to use a numerical model for the stochastic background spectrum.  With an imperfect model, there was a benefit to having the extra slope parameter that outweighed the penalty of having to explore a larger prior volume.  The two effects essentially
canceled out.  With the analytic model used in this paper, we find that the extra slope parameter does make a difference.  In Figs.~\ref{fig:bayesMLDC_NoGal} -~\ref{fig:bayesFull_LowDiv}, the model with spectral slope fitting always performs worse than the model with $m=0$.  The effect is not large and does not significantly inflate the bounds that could be placed
on a stochastic background.

\subsection{Comparison to MLDC}\label{CompareMLDC}

As a consistency check, we compare the analytic model used in this paper, to the numerical model used in Paper I.  We ran our analytic model on data containing instrument noise and a stochastic background, but no galaxy.

\begin{figure}[htbp]
   \centering
   \includegraphics[width=3.3in] {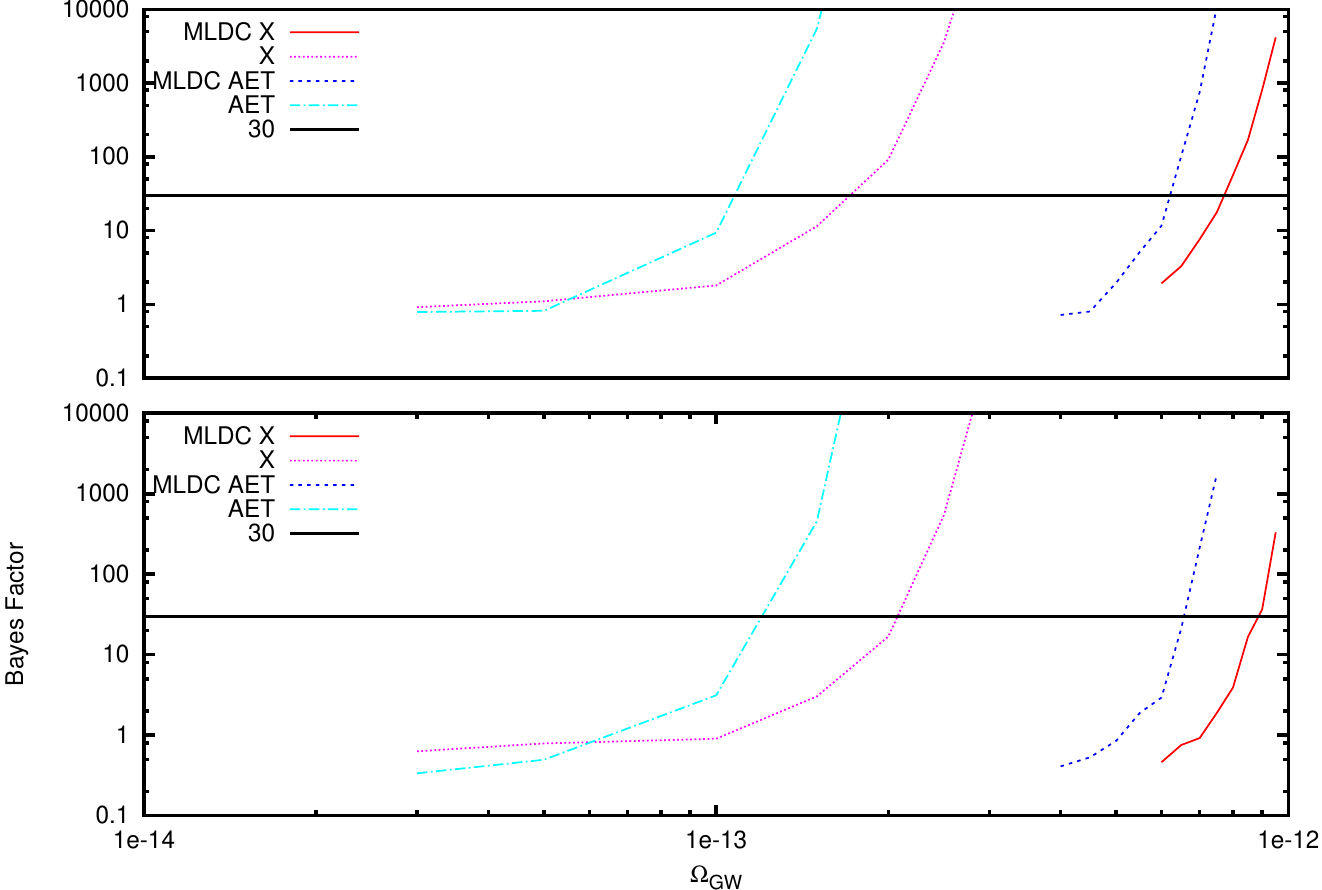} 
   \caption{A comparison of the Bayes factors for the MLDC data and for our simulated data without a galaxy.  The lower panel includes slope fitting.  A Bayes factor of 30 is considered a strong detection.    }
   \label{fig:bayesMLDC_NoGal}
\end{figure}

We find that our results are consistent with our work on the MLDC data in Paper I.   
We would expect that a 1-year data set should perform approximately $\sqrt{15}$ better than the MLDC data which is approximately 3 weeks long.  In Fig.~\ref{fig:bayesMLDC_NoGal}, we see that the new results agree very well with the MLDC results after taking into account the different observation times.  
We would expect that having a third interferometer arm would give a significant advantage in detecting an extragalactic background~\cite{Prince:2002hp, Hogan:2001jn}.  As discussed in Paper I, we don't see a large benefit with Gaussian noise, but could expect more of an advantage with more complicated noise models.

\subsection{Analysis with a Galactic foreground}
\begin{figure}[htbp]
   \centering
   \includegraphics[width=3.3in] {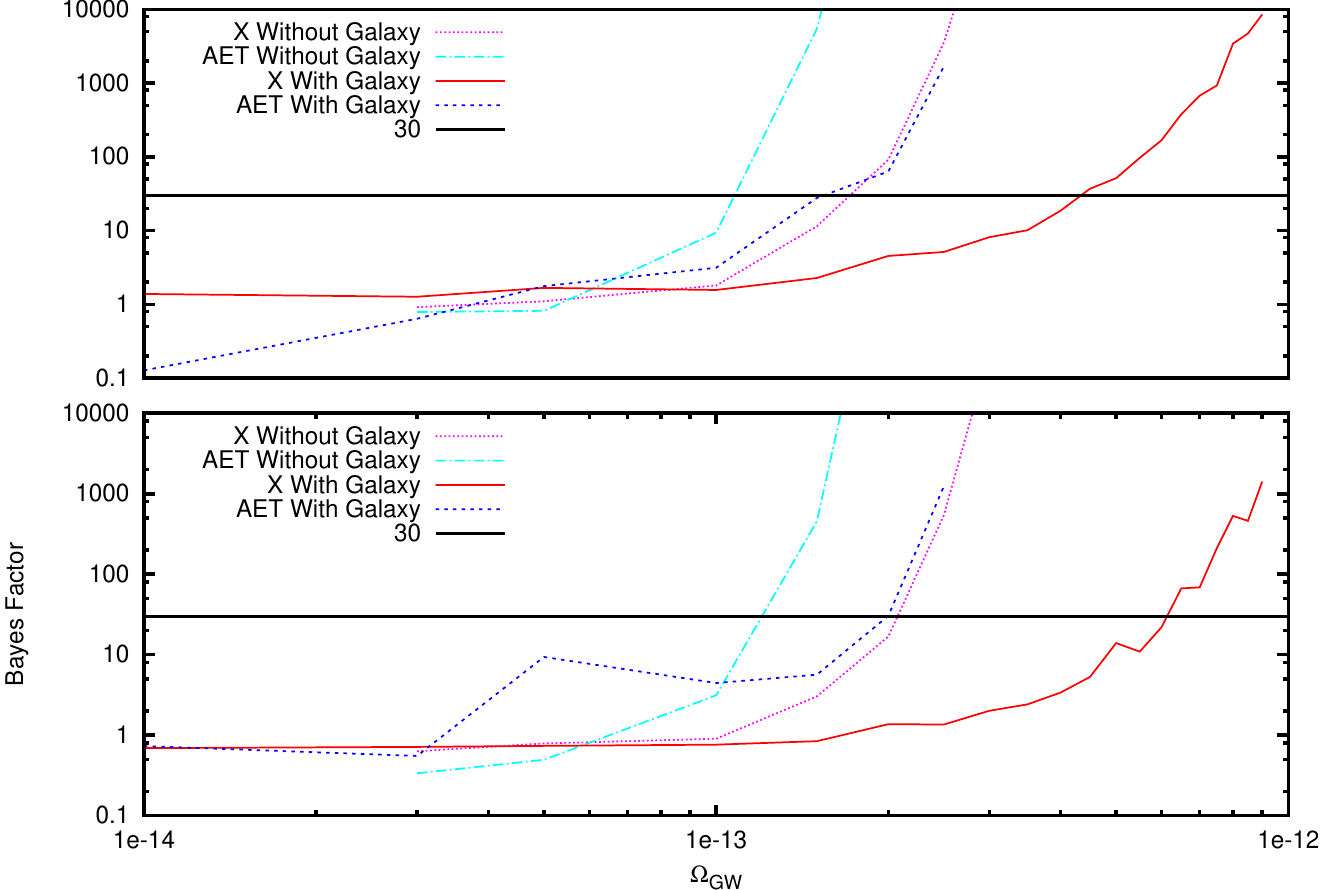} 
   \caption{A comparison of Bayes factors for data without the galaxy vs data with the galaxy included.  Including the galaxy does not significantly decrease our detection ability.  The lower panel includes slope fitting.}
   \label{fig:bayesFull_NoGal}
\end{figure}

We now compare our results for the data set with no galaxy component versus a data set that includes instrument noise, a stochastic background, and a galactic confusion foreground.  We find that our recovery of the stochastic background is not significantly diminished when including the galactic foreground.  Fig.~\ref{fig:bayesFull_NoGal}
compares the Bayes factors for the analysis without the confusion foreground to those that contain the galaxy.  The addition of the galaxy adds a small amount of uncertainty into the posterior distributions of each parameter, and we see that the detection threshold is slightly raised.  The effect is more pronounced for the X channel where the correlations are larger.  However, the Bayes factors show that the correlation between the various model components is not overwhelming and that the model is successfully distinguishing amongst the three components.

\begin{figure}[htbp]
   \centering
   \includegraphics[width=3.3in] {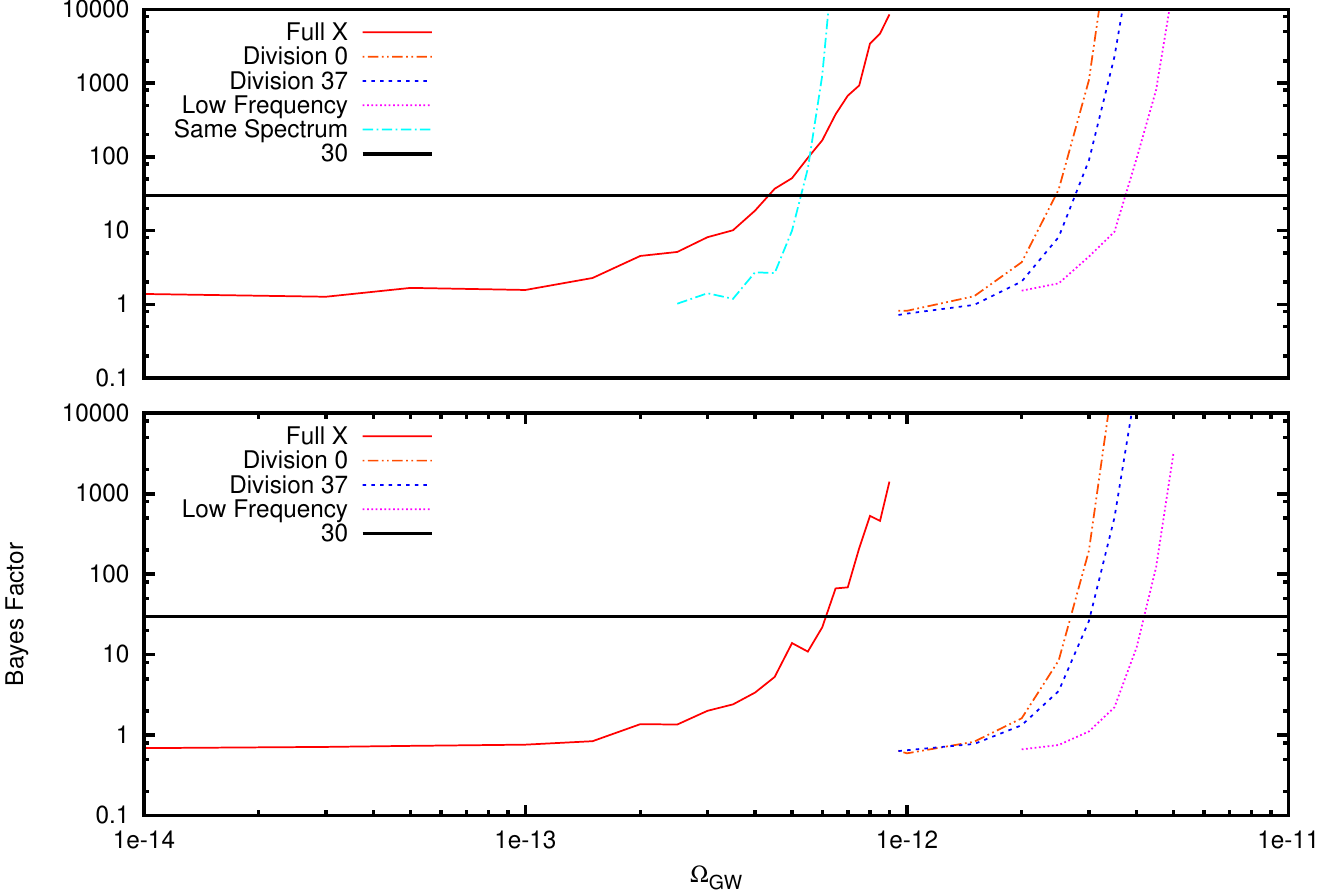} 
   \caption{A comparison of Bayes Factors for the full AET vs running on the low frequency end of the spectrum, single weeks of data, and a stochastic background with a spectral shape identical to the galaxy spectrum used in this paper.  We see that both the spectral shape and the galaxy signal modulation help separate the three model components.  The lower panel includes slope fitting.}
   \label{fig:bayesFull_LowDiv}
\end{figure}

We investigated the two effects that enable us to separate the galactic foreground from the stochastic background, namely, the shape of the spectrum and the time modulation of the signal throughout the year.  Before, we relied solely on the discriminating power of the different spectral shapes for the noise and stochastic background.  Fig.~\ref{fig:bayesFull_LowDiv} shows how both the modulation and spectral shape help.  To show that our method does not depend solely on the galaxy modulation, we ran our analysis on a single week of data.  We used the first week of the year when the foreground signal is at a minimum, and the 37th week when the foreground is at a maximum.  For one week of observation time, the foreground signal is essentially constant, and we won't get any information from the modulation.  With only one week of data, we have fewer data points and would expect to perform worse by a factor of approximately $\sqrt{50}\simeq 7$, which is what we see.  The main discriminating power in our method comes from the different spectral shapes of the various components.

Next we showed what happens if we remove the spectral information and rely on the time domain modulation alone.  We created a stochastic background that has the exact spectrum of the galaxy.  We did this by generating a separate galaxy realization and used the spectral shape from the second realization as the spectrum for the stochastic background.  The spectrum is set at different levels to determine the detection limit.  The galaxy and the background that uses the galaxy spectrum are correlated to a much greater degree, and the X-channel stochastic background would not be distinguishable without priors from the bright sources.  Additionally, the physical requirement that the X-channel power spectrum be positive for both the galaxy and stochastic background components constrains the possible values for each.   The stochastic background spectrum matching the galaxy spectrum is a worst case scenario.  In general, they will have differences that will help to better constrain the two levels.   The Bayes factors for the case of the same spectrum in Fig.~\ref{fig:bayesFull_LowDiv} should be taken as a proof of principle and not compared directly to the other Bayes factors.  The data sets with a correct background spectrum and the data set where the background spectrum matches the galaxy spectrum have very different correlations.  In the same spectrum case, the stochastic background spectrum drops below the noise levels at lower frequencies and does not correlate as much with the acceleration noise.

Lastly, we ran on the low frequency end of the spectrum (up to 3 mHz).  The acceleration noise, galaxy, and stochastic background are significant out to $\sim 5$ mHz, after which the position noise begins to dominate.  Running on the low frequency end of the spectrum gives all of the various signal components approximately equal weighting since they all extend over approximately the same number of frequency bins.  The background is recovered at a higher level, showing that we do gain by using the high frequency information to pin down the
position noise levels.  With the low frequency bins only, the position noise becomes correlated with the other components.

\section{Future Work} \label{sec:Future}
In this paper, we have outlined a very promising approach for modeling the galactic foreground and instrument noise for a space based detector so that we can look for extragalactic signals.  A stochastic background at the level of $\Omega_{GW} \sim 10^{-13}$ is very optimistic.  Standard inflation models predict that the background would be at a level of approximately $\Omega_{GW} \sim 10^{-17}$.   However, a LISA-like detector would set the best experimental bound in the milliHertz frequency regime, and may potentially uncover phase
transitions in the early Universe~\cite{Maggiore:1999vm, Giblin:2011yh, Leitao:2012tx, Caprini:2009fx, Dufaux:2012rs, Binetruy:2012ze}, or astrophysical backgrounds from EMRIs~\cite{Barack:2005aj} or extra-galactic white dwarfs.  There may be a background of extragalactic white dwarf binaries at a level of approximately $\Omega_{GW} \sim 10^{-12}$, which would be easily detected using our method~\cite{Farmer:2003pa}.  We would expect this background to be fairly isotropic, but it may be possible to see some hint of anisotropy due to the stronger signal from nearby galaxies.  Our study indicates that most of our discriminating power comes from the difference in spectral shapes.  Therefore, if the extragalactic white dwarf binary spectrum is unique, we can expect to be able to detect the background and separate it from the other stochastic signal components.   Less certain, but still an interesting possibility is a stochastic background from inspirals of massive black hole binaries~\cite{Sesana:2004gf}.

\section{Acknowledgments}
This work was supported by NASA grant NNX10AH15G. 

\bibliography{sgwb2.bib}

\end{document}